\documentclass[journal=nalefd,manuscript=letter,layout=traditional]{achemso}
\usepackage[version=3]{mhchem} 

\usepackage{bm}
\usepackage{physics}
\usepackage{graphicx}
\graphicspath{ {./figures/} }
\usepackage{caption}
\usepackage{float}
\usepackage{booktabs}
\usepackage{multirow}
\usepackage{pdflscape}
\usepackage{subcaption}
\usepackage{xcolor}
\usepackage{hyperref}
\mciteErrorOnUnknownfalse 
\hypersetup{pdfborder=0 0 0,colorlinks=true,citecolor=blue,linkcolor=blue}

\usepackage{pifont}

\def\mathcolor#1#{\@mathcolor{#1}}
\def\@mathcolor#1#2#3{%
  \protect\leavevmode
  \begingroup
    \color#1{#2}#3%
  \endgroup
}
\makeatother

\title[]{Chemical Mapping of Excitons in Halide Double Perovskites}

\author{Raisa-Ioana Biega}
\affiliation{MESA+ Institute for Nanotechnology, University of Twente, 7500 AE Enschede, The Netherlands}

\author{Yinan Chen}
\affiliation{Department of Physics,  University of Oxford, Clarendon Laboratory, Oxford,  OX1 3PU, United Kingdom}

\author{Marina R. Filip}
\email{marina.filip@physics.ox.ac.uk}
\affiliation{Department of Physics,  University of Oxford, Clarendon Laboratory, Oxford,  OX1 3PU, United Kingdom}

\author{Linn Leppert}
\email{l.leppert@utwente.nl}
\affiliation{MESA+ Institute for Nanotechnology, University of Twente, 7500 AE Enschede, The Netherlands}

\begin{document}
\begin{abstract}
Halide double perovskites are an emerging class of semiconductors with tremendous chemical and electronic diversity. While their bandstructure features can be understood from frontier-orbital models, chemical intuition for optical excitations remains incomplete. Here, we use \textit{ab initio} many-body perturbation theory within the $GW$ and the Bethe-Salpeter Equation approach to calculate excited-state properties of a representative range of Cs$_2$BB$'$Cl$_6$ double perovskites. Our calculations reveal that double perovskites with different combinations of B and B$'$ cations display a broad variety of electronic bandstructures and dielectric properties, and form excitons with binding energies ranging over several orders of magnitude. We correlate these properties with the orbital-induced anisotropy of charge-carrier effective masses and the long-range behavior of the dielectric function, by comparing with the canonical conditions of the Wannier-Mott model. Furthermore, we derive chemically intuitive rules for predicting the nature of excitons in halide double perovskites using computationally inexpensive DFT calculations.

\begin{tocentry}
    \begin{figure}[H]
        \centering
        \includegraphics[width=1\columnwidth]{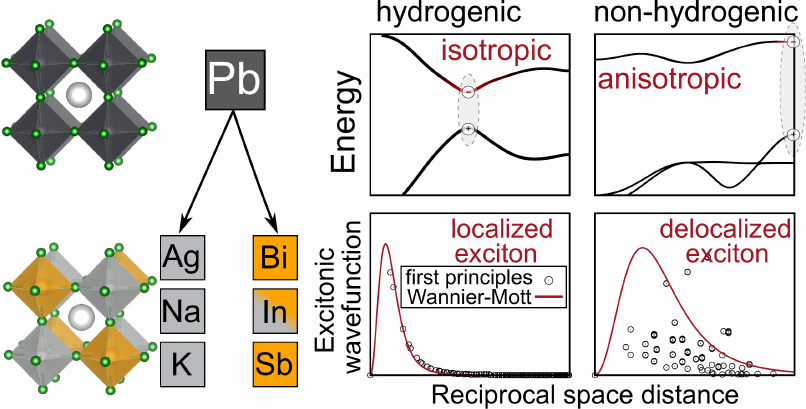}
        \label{TOC}
    \end{figure}
\end{tocentry}

\end{abstract}
Halide double perovskites, also known as elpasolites~\cite{Cross1883}, are a class of materials, with the general formula A$_2$BB$'$X$_6$, where A is a monovalent cation such as Cs$^+$, balancing the charge of corner-connected BX$_6$ and B$'$X$_6$ metal-halide octahedra. These stable, non-toxic, and earth-abundant semiconductors have showcased their potential in a range of applications, including photovoltaics~\cite{McClure2016, Slavney2016a, Greul2017, Zhao2017, Volonakis2017, Debbichi2018, Fakharuddin2019, Yang2020, Longo2020, sirtl_2d3d_2022}, X-ray detection~\cite{pan_cs2agbibr6_2017, steele_photophysical_2018}, radiation detection~\cite{Brittain1985, Pawlik1997, Biswas2012}, white light emission~\cite{Yang2018, Luo2018}, and photocatalysis \cite{muscarella_halide_2022}. This is in large part due to the tremendous chemical and structural diversity of this material class~\cite{Brandt2015, Faber2016, Volonakis2016, Filip2018}, that can be achieved by chemical substitution at the B, B$'$, and X sites~\cite{Deng2016, Jain2017, Wolf2021}.

Understanding optical excitations in halide double perovskites is crucial for designing efficient optoelectronic applications~\cite{Schade2019, Roknuzzaman2019, Longo2020, Dey2020}. In particular the binding energy of photoexcited electron-hole pairs (excitons) is a useful parameter to determine in studies of charge-carrier transport and recombination, and is thus key for device performance and design. Experimentally, exciton binding energies of halide perovskites have been extracted from optical absorption measurements either by fitting spectra using Elliott's theory~\cite{Elliott1957, Davies2018, Wright2021} or by measuring optical absorption spectra under high magnetic fields~\cite{Miyata2015a}. These methods generally assume that excitons obey the Wannier-Mott (or hydrogenic) model, which in 3D yields the following expression for the energies of the bound exciton states: $\displaystyle{E_n = -\frac{\mu}{\varepsilon^2} \cdot \frac{1}{n^2}}$ (in atomic units), where $\mu$ is the reduced effective mass, $\varepsilon$ is the dielectric constant, $n$ is the principal quantum number and with the binding energy defined as $E_B=-E_1$. The hydrogenic model has been used to understand the photophysics of a wide range of materials, from Pb-based halide perovskites~\cite{hirasawa_magnetoabsorption_1994, tanaka_comparative_2003, Miyata2015a, Galkowski2016a, Davies2018, wang_thickness-dependent_2022} to MoS$_2$ and other layered materials~\cite{chernikov_exciton_2014, selig_excitonic_2016, raja_coulomb_2017}. 
Fundamentally, the hydrogenic model relies on two main assumptions, that electronic bands must be isotropic and parabolic, and that the dielectric screening of the electron-hole interaction must be uniform (described by the constant relative permittivity)~\cite{Wannier1937}. The degree to which complex heterogeneous semiconductors abide by these tenets determines how accurate the hydrogenic picture is in describing excitons in a material, or (as we denote herein) how `{\it hydrogenic}' excitons are in a material. 

First-principles many-body perturbation theory within the $GW$ approximation~\cite{Fetter1971} and the Bethe-Salpeter Equation~\cite{Rohlfing1998, Rohlfing2000} (BSE) approach has played a particularly important role in quantitatively predicting the electronic and excited-state structure of halide perovskites. In particular, comparison of $GW$+BSE calculations with the Wannier-Mott model has demonstrated the hydrogenic nature of excitons in Pb-based halide perovskites~\cite{Bokdam2016, Filip2021} and in the double perovskite Cs$_2$AgInCl$_6$ \cite{Luo2018, Jain2022}. In contrast, we and others recently showed that the double perovskite family Cs$_2$AgBX$_6$ (B=Bi, Sb and X=Br, Cl)~\cite{McClure2016, Slavney2016a, Volonakis2016} exhibits resonant excitons with binding energies between 170 and 450\,meV which are strongly localized, with fine-structure features that differ from those computed using the hydrogenic model~\cite{Palummo2020, Biega2021}. We assigned the non-hydrogenic character of excitons in these halide double perovskites to their chemical heterogeneity giving rise to an anisotropic electronic structure and dielectric screening~\cite{Biega2021}. For other halide double perovskites, optoelectronic properties and exciton binding energies were shown to vary significantly too \cite{Luo2022, Yu2023, Adhikari2023}. The picture that emerges from these reports suggests a rich landscape of excitons in halide double perovskites and calls for a systematic mapping of this landscape using first-principles calculations.

In this letter, we use the $GW$+BSE approach to develop a holistic understanding of how the electronic structure of the alternating B- and B$'$-site cations influences the nature of excitons in halide double perovskites. By studying a representative set of halide double perovskites Cs$_2$BB$'$Cl$_6$, we show that exciton binding energies can be tuned by several orders of magnitude through chemical substitution at the B and B$'$ sites. Furthermore, we demonstrate that direct band gap halide double perovskites with isotropic, parabolic band edges and small local field effects in their dielectric screening, feature hydrogenic excitons similar to their Pb-based single perovskite congeners. However, the absorption spectra of these materials depend considerably on the symmetry of the band edges, and can deviate significantly from expectations prescribed by canonical models. Among the heterogeneous double perovskites we study systematically, we find that some (but not all) exhibit an exciton fine structure which is well described by the hydrogenic model. However, the extent to which excitons present as non-hydrogenic depends strongly on the electronic structure of the alternating B and B$'$ metals. 

\begin{figure}[htb]
    \centering
    \includegraphics[width=0.65\textwidth]{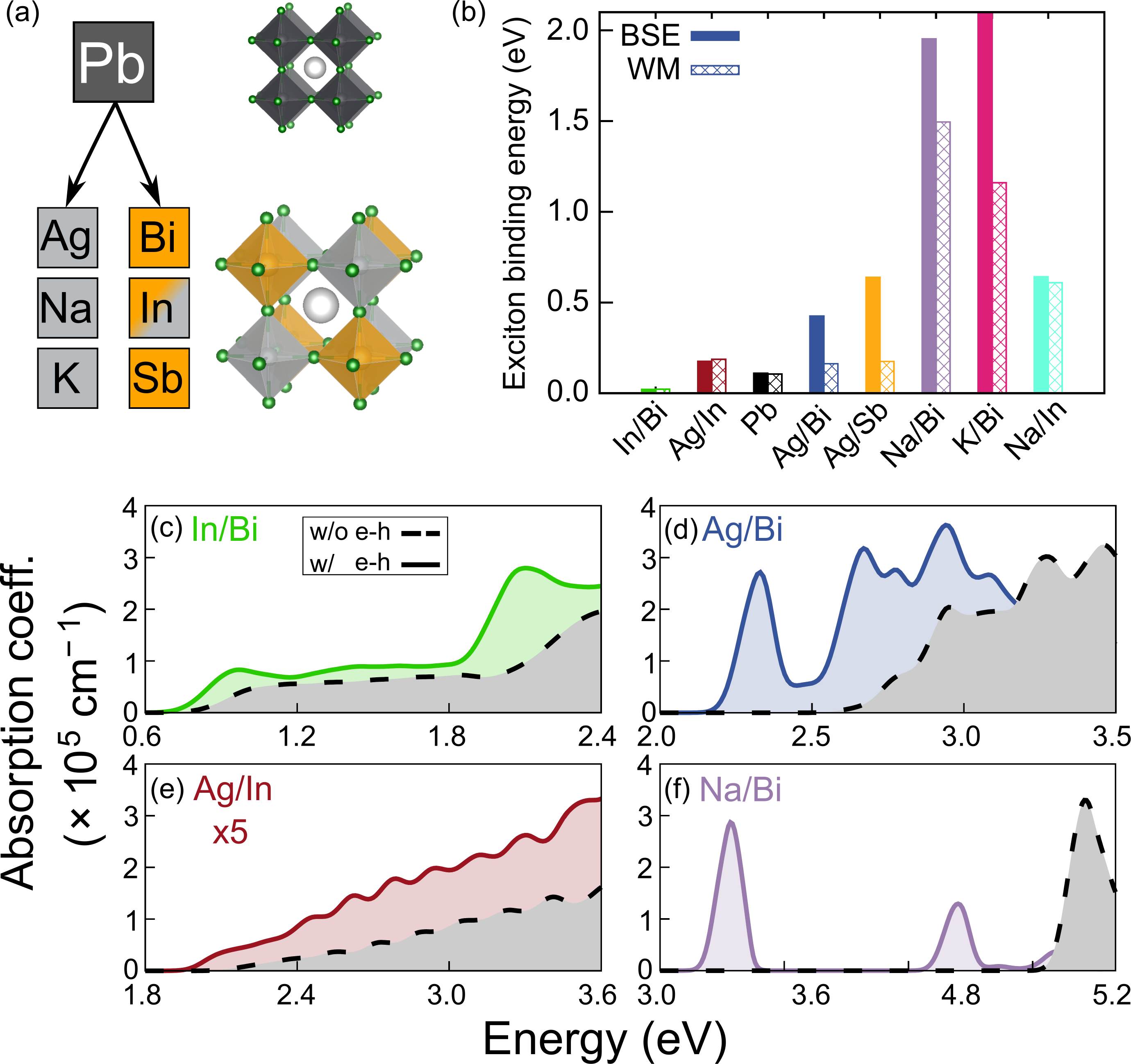}
    \caption{(a) Overview of materials studied. In the cubic structures of single and double perovskites, X sites (Cl$^-$) are green, A sites (Cs$^+$) are white, dark gray corresponds to Pb, light-grey to the B and orange to the B$'$ site. (b) Exciton binding energy computed from first principles (solid bars), and estimated based on the Wannier-Mott fine structure, as described in the text (patterned bars). Materials appear in order of their QP band gap with the lowest band-gap material on the left. Linear optical absorption spectra, calculated using the independent-particle approximation (black dashed line), and the $G_0W_0$+BSE approach (solid colored line) for (c) \textbf{In/Bi}, (d) \textbf{Ag/Bi}, (e) \textbf{Ag/In} and (f) \textbf{Na/Bi}.}
    \label{fig:bind-en-abs-spectra}
\end{figure}

Recently, Ref.~\citenum{Filip2021} showed that the fully inorganic Pb-based halide perovskites feature hydrogenic excitons. Here we use the cubic phase of CsPbCl$_3$ (referred to as \textbf{Pb} in the following) as a prototypical case of a direct band gap single perovskite, which we compare to seven representative cubic double perovskites A$_2$BB$'$X$_6$ with A = Cs$^+$ and X = Cl$^-$ (denoted by \textbf{B/B$'$} in the following). Our goal is to identify how the electronic structure of the B- and B$'$-site cations determines the hydrogenic nature of excitons in halide double perovskites. To this end, we explore double perovskites featuring metals from across the periodic table (Figure~\ref{fig:bind-en-abs-spectra}(a)): \textbf{In/Bi}, which is isoelectronic to \textbf{Pb}; \textbf{Ag/In} and \textbf{Na/In} which feature a direct band gap, and large band dispersion, but a distinctly different band edge orbital character than \textbf{In/Bi}; and \textbf{Ag/Bi}, \textbf{Ag/Sb}, \textbf{Na/Bi} and \textbf{K/Bi} with an indirect band gap and low-dispersion band edges. With the exception of \textbf{In/Bi}, these double perovskites have all been synthesized and experimentally characterized~\cite{Slavney2016a, McClure2016, Tran2017, Steele2018, Bartesaghi2018a, Luo2018, Dahl2019, Majher2019, Noculak2020, Gray2020}. The Na- and K-based compounds have experimentally been studied as favorable host structures for luminescent centers such as Mn$^{2+}$ and Sb$^{3+}$~\cite{Majher2019, Noculak2020, Gray2020}. However, to the best of our knowledge, we are the first to perform state-of-the-art $GW$+BSE calculations for these materials and report their exciton binding energies.

We start by calculating the quasiparticle (QP) band structure, absorption spectra and exciton binding energies of all eight materials using the $GW$+BSE approach as implemented in the \textsc{BerkeleyGW} code \cite{Deslippe2012, barker_spinor_2022} (see the Supporting Information (SI) and Figures S1 -- S3 for further computational details and convergence studies). Figure~\ref{fig:bind-en-abs-spectra}(b) shows the exciton binding energy from first-principles calculations (BSE) and estimated according to the Wannier-Mott fine structure (WM). BSE exciton binding energies ($E_\text{BSE}$) are calculated as the difference between the energy of the first excited state and the direct QP band gap. To avoid systematic errors in the calculation of Wannier-Mott energy (see SI for details), here, and unless otherwise noted, we quantify agreement with the Wannier-Mott model using the excitonic fine structure (excited state energy levels) $\displaystyle{E^\text{fs, WM} = \frac{4}{3}\cdot(E_\text{2s}-E_\text{1s})}$, i.e., from the difference between the $G_0W_0$+BSE excitation energies of the $1s$ ($E_\text{1s}$) and $2s$ ($E_\text{2s}$) states, respectively (see SI and Figure~S4 for exciton fine structures and assignment of the $1s$ and $2s$ states). Table~\ref{tab:exc-bind-en} reports the QP band gaps and exciton binding energies of all eight materials.

Figure~\ref{fig:bind-en-abs-spectra}(b) and Table~\ref{tab:exc-bind-en} allow for several observations: First, our selected double perovskites span a wide range of QP band gaps between $\sim$1 and 5\,eV, which are inversely proportional to their dielectric constants $\varepsilon_{\infty}$ (see Figure~S5). The exciton binding energies of these compounds differ by several orders of magnitude, with $E_\text{BSE}$ ranging from 16\,meV (\textbf{In/Bi}) to $\sim$2\,eV (\textbf{K/Bi}). However, as shown in Figure~S6, $E_\text{BSE}$ does not scale linearly with $1/\varepsilon_{\infty}^2$, suggesting that the Wannier-Mott model performs poorly for a subset of double perovskites. Indeed, the first-principles exciton binding energies of \textbf{Ag/Bi}, \textbf{Ag/Sb}, \textbf{Na/Bi}, and \textbf{K/Bi} deviate by several hundred meV from the Wannier-Mott fine structure. In contrast, and despite their seemingly similar degree of chemical heterogeneity, \textbf{Ag/In}, \textbf{In/Bi}, and \textbf{Na/In} feature hydrogenic excitons, similar to the single perovskite \textbf{Pb}~\cite{Filip2021}. We therefore separate the studied double perovskites in two groups: materials with \textit{hydrogenic} (\textbf{Pb}, \textbf{Bi/In}, \textbf{Ag/In}, \textbf{Na/In}) and materials with \textit{non-hydrogenic} (\textbf{Ag/Bi}, \textbf{Ag/Sb}, \textbf{Na/Bi}, \textbf{K/Bi}) exciton fine structures. Notably, we find that $\Delta_\text{WM} = E_{\text{BSE}} - E^\text{fs, WM}$, does not necessarily increase with increasing exciton binding energy. In other words, the magnitude of the exciton binding energy does not necessarily explain the observed non-hydrogenic fine structure. For example, \textbf{Na/In} features a hydrogenic $1s$ exciton with a very high binding energy of $\sim$600\,meV.

Not only the exciton binding energies, but also the linear optical absorption spectra of these eight materials differ significantly, as shown for representative double perovskites with \textit{hydrogenic} and \textit{non-hydrogenic} excitons in Figure~\ref{fig:bind-en-abs-spectra}(c), (e) and (d), (f), respectively (see also Figure~S7). The absorption spectrum of \textbf{In/Bi} exhibits a distinct excitonic feature, similar to the isoelectronic \textbf{Pb}. In agreement with previous results~\cite{Luo2018}, we observe that \textbf{Ag/In} and \textbf{Na/In} have a weak absorption onset and do not exhibit a signature excitonic peak. The absorption coefficient is also one order of magnitude lower than that of the other materials. This is in line with the dipole-forbidden transitions between the valence and conduction band edges~\cite{Meng2017} of these materials. In contrast, all four materials with \textit{non-hydrogenic} excitons feature one or several distinct excitonic peaks at the onset of absorption. 

\begin{table*}[ht]
\small
\centering
\begin{tabular}{@{}cccccc@{}}
\textit{hydrogenic}\\
\toprule
    \textbf{System} & \textbf{QP direct} & \textbf{Static dielectric} & \multicolumn{3}{c}{\textbf{Exciton binding energy (eV)}}\\
    \textbf{B/B$'$} & \textbf{gap (eV)} & \textbf{constant} & \multicolumn{2}{c}{\textbf{BSE}} & \textbf{Wannier-Mott}\\
    & $E_\text{gap}^{G_0W_0}$ & $\varepsilon_\infty$ & $E^\text{BSE}_\text{dark}$ & $E^\text{BSE}_\text{bright}$ & $E^\text{fs, WM}$\\\midrule
    \textbf{Pb}    & 2.25 & 3.67 & 0.105 & 0.104 & 0.103 \\
    \textbf{In/Bi} & 0.90 & 5.63 & 0.021 & 0.020 & 0.018 \\
    \textbf{Ag/In} & 2.09 & 3.75 & 0.176 & 0.049 & 0.170 \\
    \textbf{Na/In} & 5.52 & 2.78 & 0.642 & 0.135 & 0.605 \\\midrule
\bottomrule\\
\textit{non-hydrogenic}\\
\toprule
    \textbf{System} & \textbf{QP direct} & \textbf{Static dielectric} & \multicolumn{3}{c}{\textbf{Exciton binding energy (eV)}}\\
    \textbf{B/B$'$} & \textbf{gap (eV)} & \textbf{constant} & \multicolumn{2}{c}{\textbf{BSE}} & \textbf{Wannier-Mott}\\
    & $E_\text{gap}^{G_0W_0}$ & $\varepsilon_\infty$ & $E^\text{BSE}_\text{dark}$ & $E^\text{BSE}_\text{bright}$ & $E^\text{fs, WM}$\\\midrule
    \textbf{Ag/Bi} & 2.64 & 4.49 & 0.426 & 0.329 & 0.206 \\
    \textbf{Ag/Sb} & 3.20 & 4.63 & 0.639 & 0.549 & 0.302 \\
    \textbf{Na/Bi} & 4.93 & 3.09 & 1.953 & 1.611 & 0.553 \\
    \textbf{ K/Bi} & 4.99 & 2.73 & 2.091 & 1.725 & 0.487 \\\midrule
\bottomrule
\end{tabular}
\caption{$G_0W_0$@PBE lowest direct band gap, static dielectric constant as computed within the random phase approximation, $E^\text{BSE}$ of the first dark and first bright excited state and $E^\text{fs, WM}$.}
\label{tab:exc-bind-en}
\end{table*}

\begin{figure*}[h!]
    \centering
    \includegraphics[width=1.\textwidth]{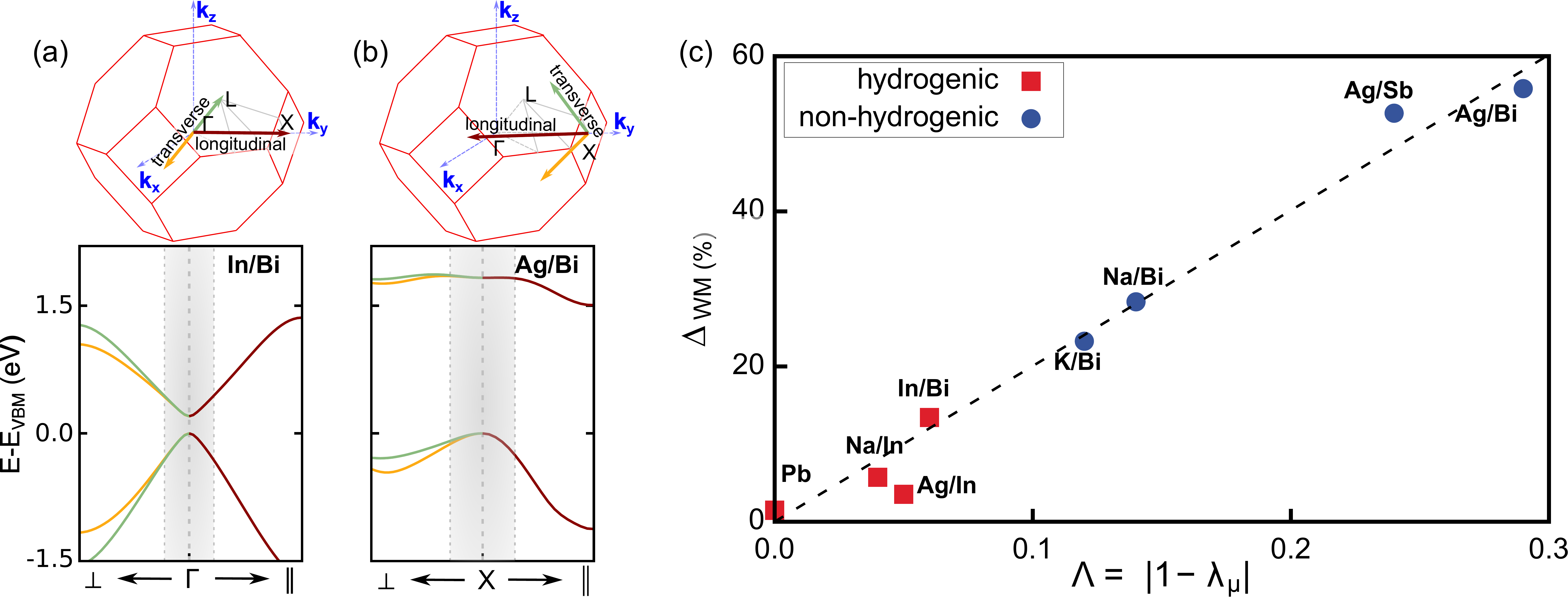}
    \caption{Graphical representation of the Brillouin zone and the DFT-PBE+SOC conduction and valence bands of (a) \textbf{In/Bi} and (b) \textbf{Ag/Bi} along the transversal (X/$\Gamma{}\to{}\perp$) and longitudinal (X/$\Gamma{}\to{}\parallel$) directions. The shaded area corresponds to the \textbf{k}-interval used to compute the effective masses. (c) Variation of the relative deviation $\Delta_\text{WM}$ of the Wannier-Mott fine structure with respect to the first-principles (BSE) result as a function of the effective mass anisotropy $\Lambda=\abs{1-\lambda_\mu}$.}
    \label{fig:3d-bands-anisotropy}
\end{figure*}

Having established these subsets of materials with \textit{hydrogenic} and \textit{non-hydrogenic} excitons, we continue by probing to which degree the main assumptions of the Wannier-Mott model -- isotropic, parabolic band edges and a uniform, isotropic dielectric constant -- are fulfilled for these materials. We start by calculating the effective electron and hole masses at the high-symmetry point in the Brillouin zone of the lowest-energy direct transition (Table~S3 and Figures~S8 and S9 for DFT and $G_0W_0$ band gaps and bandstructures) along the principal axes of the effective mass tensor, which can be identified as longitudinal and transverse effective masses (Table~\ref{tab:deviation-exc-bind-en}), similar to $fcc$ semiconductors such as Si and GaAs~\cite{Yu2001} (see SI). We show the valence and conduction band edges along those directions in Figure~\ref{fig:3d-bands-anisotropy}(a) and (b), for \textbf{In/Bi} and \textbf{Ag/Bi}, representative for double perovskites with \textit{hydrogenic} and \textit{non-hydrogenic} excitons, respectively. Around the high symmetry point of the lowest-energy direct transition -- $\Gamma$ [0, 0, 0] for \textbf{In/Bi} and X [0, 1, 0]$2\pi/a$ for \textbf{Ag/Bi} -- the band edges are isotropic for \textbf{In/Bi} and highly anisotropic for \textbf{Ag/Bi}, with different curvatures in the longitudinal and transverse directions. 

Further analysis reveals that the effective mass anisotropy factor $\displaystyle{\lambda_\mu=\left(\frac{\mu_\perp}{\mu_\parallel}\right)^{1/3}}$ is close to 1 for the double perovskites with \textit{hydrogenic} excitons and exactly 1 for \textbf{Pb}. In Figure~\ref{fig:3d-bands-anisotropy}(c) we show $\Delta_{\text{WM}}$ as a function of the quantity $\Lambda = \abs{1-\lambda_\mu}$, where $\Lambda=0$ corresponds to a fully isotropic material (e.g., \textbf{Pb}). This analysis shows that the relative deviation from the Wannier-Mott model scales almost linearly with the degree of anisotropy. The materials with \textit{hydrogenic} excitons (red squares) are mostly isotropic and feature a deviation of no more than 14\,\% from the Wannier-Mott model. In contrast, the double perovskites with \textit{non-hydrogenic} excitons (blue dots) show a significantly higher degree of anisotropy and a large deviation from the Wannier-Mott model. We note that accounting for the effective mass anisotropy in the Wannier-Mott model following Ref.~\citenum{Schindlmayr1997} reduces $\Delta_\text{WM}$ but does not fully account for the observed deviations (Table~S4).

\begin{figure}[htb]
    \centering
    \includegraphics[width=0.9\textwidth]{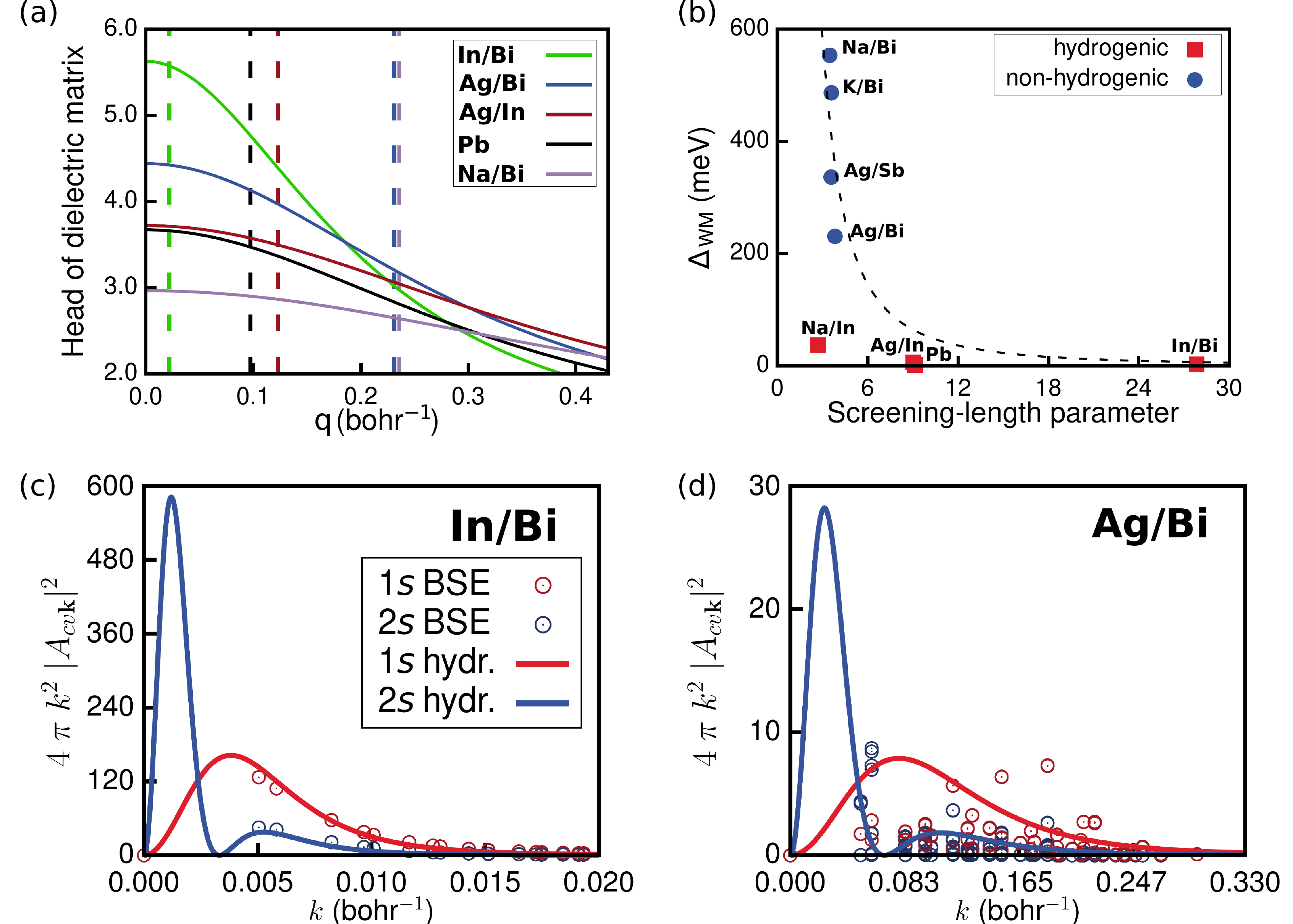}
    \caption{(a) Model dielectric function (as described in the text) in reciprocal space for \textbf{In/Bi} (green), \textbf{Ag/Bi} (blue), \textbf{Ag/In} (red), \textbf{Pb} (black), and \textbf{Na/Bi} (purple). The corresponding colored dashed line shows the exciton extent in \textbf{k}-space ($k_\text{x}$) as defined in the text. (b) Absolute variation of $\Delta_\text{hydr.}$ as a function of the screening-length parameter $\displaystyle{l=\frac{q_\text{TF}}{k_\text{x}}}$. Exciton radial probability density in reciprocal space as computed from $G_0W_0$+BSE (empty disks) and as predicted by the Wannier-Mott model (solid lines) for $1s$ (red) and $2s$ (blue) states for (c) \textbf{In/Bi} and (d) \textbf{Ag/Bi}.}
    \label{fig:eps-kspace-dev-screening}
\end{figure}

Next, we probe the uniformity and isotropy of the dielectric screening by computing linear absorption spectra and exciton binding energies assuming uniform dielectric screening (see SI), and find that they change significantly only for those perovskites in which excitons do not display hydrogenic behavior (Figure~S7 and Table~S6). We then analyze the spatial dependence of the head, i.e., the $\mathbf{G}=\mathbf{G'}=0$ component, of the dielectric function. Figure~\ref{fig:eps-kspace-dev-screening}(a) and Figure~S10 show that we can fit this spatial dependence with the model dielectric function, $\displaystyle{\Re[\epsilon(q,0)] = 1+\left[ \left(\varepsilon_\infty-1\right)^{-1}+\alpha\left(\frac{q}{q_{TF}}\right)^2 \right] ^{-1}}$, with $\alpha$ and the Thomas-Fermi wave vector, $\displaystyle{q_{TF}}$ as fitting parameters tabulated in Table~S6 and $\varepsilon_\infty$ the RPA dielectric constant \cite{Cappellini1993}. 

We then compare the length scale of dielectric screening  -- quantified by $q_{TF}$ -- with the length scale of the excitonic wave function of the first excited state $k_x$ in reciprocal space, by calculating the screening length parameter $\displaystyle{l=\frac{q_{TF}}{k_x}}$ (Table~\ref{tab:deviation-exc-bind-en}). For this, we define $k_x$ such that it includes 99\,\% of the exciton probability density of the first excited state. Figure~\ref{fig:eps-kspace-dev-screening}(b) shows that $\Delta_\text{WM}$ decreases as $1/l^2$. Perovskites with \textit{hydrogenic} excitons feature large screening-length parameters ($l \geq 11$), corresponding to excitons highly localized in reciprocal space, for which the dielectric screening can be assumed to be uniform and constant. An outlier is \textbf{Na/In} with $l=2.71$, closer to the materials with \textit{non-hydrogenic} excitons. We attribute this to the slow variation of the dielectric function in reciprocal space in this large band-gap compound (Figure S10). In contrast, for the subset of materials with \textit{non-hydrogenic} excitons, the variation of the dielectric constant is significant on the length scale of their excitons, strongly delocalized in reciprocal space. Deviations from the Wannier-Mott model in these materials are also apparent by comparing the excitonic wave functions computed with G$_0$W$_0$+BSE with the radial probability density of hydrogenic $1s$ and $2s$ excitonic wavefunctions, shown in Figure~\ref{fig:eps-kspace-dev-screening}(c) and (d) for \textbf{In/Bi} and \textbf{Ag/Bi}, respectively (see Figure~S11 for the other materials).

\begin{table*}[htb]
\centering
\small
    \begin{tabular}{@{}cccccc@{}}
    \textit{hydrogenic}\\
    \toprule
    \textbf{System} &  \multicolumn{2}{c}{\textbf{Reduced mass}} & \textbf{Screening-length} & \multicolumn{2}{c}{\textbf{Deviation from}}\\
    \textbf{B/B$'$} & \multicolumn{2}{c}{\textbf{and anisotropy}} & \textbf{parameter} & \multicolumn{2}{c}{\textbf{Wannier-Mott}}\\
    & $\mu$ & $\lambda_\mu$ & $\displaystyle{l=\frac{q_\text{TF}}{k_\text{x}}}$ & $\Delta_\text{hydr.}\text{ (\%)}$ & $\Delta_\text{hydr.}\text{ (meV)}$ \\ \midrule
    \textbf{Pb}    & 0.105 & 1.00 &  9.16 &  1.48 &   1.54 \\
    \textbf{In/Bi} & 0.052 & 0.94 & 27.84 & 13.43 &   2.77\\
    \textbf{Ag/In} & 0.191 & 1.05 &  9.02 &  3.54 &   6.23\\
    \textbf{Na/In} & 0.345 & 0.96 &  2.71 &  5.73 &  36.74\\ \midrule
    \bottomrule\\
    \textit{non-hydrogenic}\\
    \toprule
    \textbf{System} &  \multicolumn{2}{c}{\textbf{Reduced mass}} & \textbf{Screening-length} & \multicolumn{2}{c}{\textbf{Deviation from}}\\
    \textbf{B/B$'$} & \multicolumn{2}{c}{\textbf{and anisotropy}} & \textbf{parameter} & \multicolumn{2}{c}{\textbf{Wannier-Mott}}\\
    & $\mu$ & $\lambda_\mu$ & $\displaystyle{l=\frac{q_\text{TF}}{k_\text{x}}}$ & $\Delta_\text{hydr.}\text{ (\%)}$ & $\Delta_\text{hydr.}\text{ (meV)}$ \\ \midrule
    \textbf{Ag/Bi} & 0.241 & 1.29 & 3.82 & 55.82 & 230.83\\
    \textbf{Ag/Sb} & 0.250 & 1.23 & 3.57 & 52.66 & 336.38\\
    \textbf{Na/Bi} & 0.257 & 0.86 & 3.47 & 28.34 & 553.39\\
    \textbf{ K/Bi} & 0.636 & 0.88 & 3.58 & 23.27 & 486.63\\ \midrule
    \bottomrule
\end{tabular}
\caption{$G_0W_0$@PBE reduced effective mass $\mu$ (in units of the electron rest mass $m_0$), anisotropy factor $\lambda_\mu$, static dielectric constant (within the random phase approximation), screening-length parameter and relative and absolute deviation of the Wannier-Mott exciton binding energy with respect to the first-principles (BSE) result.}
\label{tab:deviation-exc-bind-en}
\end{table*}

Finally, we return to our goal of mapping the character of excitons in halide double perovskites to the band edge electronic structure obtained from computationally inexpensive DFT calculations (instead of a full solution of the BSE). We note that the effective mass anisotropy can be approximated using DFT effective masses (see Table~S3). Chemical intuition for the reciprocal space location, parity, and dispersion of the VBM and CBM can be obtained based on knowledge of the metal-orbital character of the band edges of halide perovskites, as shown in Ref.~\citenum{Slavney2019} using linear combinations of atomic orbitals and symmetry arguments. Table~\ref{tab:exciton-prediction} shows the calculated B-site orbital character and high-symmetry \textbf{k}-point of the VBM and CBM of all eight perovskites, in agreement with previous predictions~\cite{Slavney2019}, and lists the B-site orbital character at the lowest direct transition from which excitons are derived. We observe that in all materials with \textit{non-hydrogenic} excitons, the B-site orbital contributions to the band edges lead to an indirect band gap. The lowest direct transition in these materials occurs at the Brillouin zone boundaries (X in \textbf{Ag/Bi} and \textbf{Ag/Sb}, and L in \textbf{K/Bi} and \textbf{Na/Bi}). Furthermore, in all those materials the conduction band edge at the lowest direct transition is relatively flat along the transverse directions which is a consequence of small B and B$'$ site orbital overlap. In \textbf{Ag/Bi} and \textbf{Ag/Sb}, there is no Ag~$s$ character at the X points due to orbital symmetry. In \textbf{Na/Bi} and \textbf{K/Bi}, Na~$s$ and K~$s$ contribute to the conduction band edge at L but their overlap with the neighboring Bi~$p$ orbitals is small leading to high effective masses in the transverse direction.

In materials with \textit{hydrogenic} excitons there is a symmetry match of the B- and B$'$- site orbitals leading to a direct band gap at $\Gamma$ with isotropic effective masses. Note that a direct gap can also arise in a double perovskite in which only one or none of the metal sites contributes to the band edges \cite{Slavney2019}. In \textbf{Na/In}, the valence band edge is rather flat because it does not feature metal-site orbital contributions. However, the conduction band minimum -- with contributions from Na~$s$ and In~$s$ -- is at $\Gamma$, disperse and isotropic.

Our predictions can also be extended to double perovskites with one B-site only, such as the vacancy-ordered perovskites with chemical formula Cs$_2$BX$_6$. $GW$+BSE calculations by Cucco \textit{et al.} for Cs$_2$TeBr$_6$ and Kavanagh \textit{et al.} for Cs$_2$TiX$_6$ (X=I, Br, Cl) indicate excitons highly localized in real space and ill-described by the Wannier-Mott model for these materials~\cite{Kavanagh2022, Cucco2023}. Cs$_2$SnX$_6$, on the other hand, features \textit{hydrogenic} excitons, as reported in both Ref.~\citenum{Kavanagh2022} and \citenum{Cucco2023}. Using Cs$_2$TeBr$_6$, Cs$_2$TiBr$_6$, and Cs$_2$SnBr$_6$ as examples, we calculated the orbital character of the band edges and effective mass anisotropy using DFT (see Table~S7). The band structure of Cs$_2$TeBr$_6$ is reminiscent of that of \textbf{Ag/Bi} and \textbf{Ag/Sb} with Te~$6s$ and $6p$ contributions leading to an indirect band gap. The lowest direct transition is at L, where the conduction band is derived from Te~$p$ orbitals alone and therefore anisotropic. Cs$_2$TiBr$_6$ is another indirect-gap semiconductor. Its relatively flat and highly anisotropic conduction band is derived from localized Ti~$d$ states and features weak Ti~$d$–X~$p$ mixing. Cs$_2$SnBr$_6$ on the other hand is comparable to \textbf{Na/In} with a direct band gap at $\Gamma$ and an isotropic conduction band derived from Sn~$s$ orbitals (Table~\ref{tab:exciton-prediction}). Thus, for the systems analyzed here, an effective mass anisotropy factor larger than 0.1 appears to be sufficient to predict a \textit{non-hydrogenic} excitonic fine structure. However, as shown in Tables~S4 and S5, effective mass anisotropy is only one contributing factor, alongside the non-uniformity of the dielectric function, in particular in materials with larger dielectric constants.

\begin{table*}[htb]
\centering
\resizebox*{1.\linewidth}{!}{
\begin{tabular}{@{}c|ccc|ccc|cc|c@{}}
    \toprule
    \multirow{2}{*}{\textbf{System}} & \multicolumn{3}{c}{\textbf{Band gap}} & \multicolumn{3}{c}{\textbf{Lowest direct trans.}} & \multicolumn{2}{c}{\textbf{Anisotropy}} & \multirow{2}{*}{\textbf{Exciton}}\\
    & valence & conduction & \textbf{k}-point & valence & conduction & \textbf{k}-point & pred. & calc. & \\
    \midrule
    \multirow{2}{*}{\textbf{Ag/Bi}} & Ag $4d_{z^2}$ & Ag $5s$ & \multirow{2}{*}{Indirect X$\to$L} & Ag $4d_{z^2}$ & \textit{Null} 
    & \multirow{2}{*}{X} & \multirow{2}{*}{high $\Lambda$} & \multirow{2}{*}{high $\Lambda$} & \multirow{2}{*}{non-hydr.}\\
    & Bi $6s$ & Bi $6p$ & & Bi $6s$ & Bi $6p$ & & &\\
    \rule{0pt}{4ex}
    \multirow{2}{*}{\textbf{Ag/Sb}} & Ag $4d_{z^2}$ & Ag $5s$ &	\multirow{2}{*}{Indirect X$\to$L} & Ag $4d_{z^2}$ & \textit{Null} 
    & \multirow{2}{*}{X} &\multirow{2}{*}{high $\Lambda$} & \multirow{2}{*}{high $\Lambda$} & \multirow{2}{*}{non-hydr.}\\
    & Sb $5s$ & Sb $5p$ & & Sb $5s$ & Sb $5p$ & & &\\
    \rule{0pt}{4ex}
    \multirow{2}{*}{\textbf{K/Bi}} & \textit{Null}
    & K $4s$ & Indirect L$\to\Gamma$ & \textit{Null} 
    & K $4s$ & \multirow{2}{*}{L} & \multirow{2}{*}{high $\Lambda$} & \multirow{2}{*}{high $\Lambda$} & \multirow{2}{*}{non-hydr.}\\
    & Bi $6s$ & Bi $6p$ & Direct $\Gamma$ & Bi $6s$ & Bi $6p$ & & &\\
    \rule{0pt}{4ex}
    \multirow{2}{*}{\textbf{Na/Bi}} & \textit{Null} 
    & Bi $6p$ & \multirow{2}{*}{Indirect X$\to\Gamma$} & \textit{Null} 
    & Na $3s$ & \multirow{2}{*}{L} & \multirow{2}{*}{high $\Lambda$} & \multirow{2}{*}{high $\Lambda$} & \multirow{2}{*}{non-hydr.}\\
    &  Bi $6s$ & Na $3s$ & & Bi $6s$ & Bi $6p$ & & &\\
    \rule{0pt}{4ex}
    \multirow{2}{*}{\textbf{Ag/In}} & Ag $4d_{z^2/x^2-y^2}$ & Ag $5s$ & \multirow{2}{*}{Direct $\Gamma$} & Ag $4d_{z^2/x^2-y^2}$ & Ag $5s$ & \multirow{2}{*}{$\Gamma$} & \multirow{2}{*}{low $\Lambda$} & \multirow{2}{*}{low $\Lambda$} & \multirow{2}{*}{hydr.}\\
    & \textit{Null} 
    & In $5s$ & & \textit{Null} 
    & In $5s$ & & &\\
    \rule{0pt}{4ex}
    \multirow{2}{*}{\textbf{Na/In}} & \textit{Null} 
    & Na $3s$ & \multirow{2}{*}{Direct $\Gamma$} & \textit{Null} 
    & Na $3s$ & \multirow{2}{*}{$\Gamma$} & \multirow{2}{*}{low $\Lambda$} & \multirow{2}{*}{low $\Lambda$} & \multirow{2}{*}{hydr.}\\
    & \textit{Null} 
    & In $5s$ & & \textit{Null} & In $5s$ & & &\\
    \rule{0pt}{4ex}
    \multirow{2}{*}{\textbf{In/Bi}} & In $5s$ & In $5p$ & \multirow{2}{*}{Direct $\Gamma$} & In $5s$ & In $5p$ & \multirow{2}{*}{$\Gamma$} & \multirow{2}{*}{low $\Lambda$} & \multirow{2}{*}{low $\Lambda$} & \multirow{2}{*}{hydr.}\\
    & Bi $6s$ & Bi $6p$ & & Bi $6s$ & Bi $6p$ & & &\\
    \rule{0pt}{4ex}
    \textbf{Pb} & Pb $6s$ & Pb $6p$ & Direct R & Pb $6s$ & Pb $6p$ & R & no $\Lambda$ & no $\Lambda$ & hydr.\\
    \midrule
    \rule{0pt}{4ex}
    \textbf{Cs$_2$TeBr$_6$} & Te $6s$ & Te $6p$ & Indirect X$\to$L & Te $6s$ & Te $6p$ & L & high $\Lambda$ & high $\Lambda$ & non-hydr.\\
    \rule{0pt}{4ex}
   \textbf{Cs$_2$TiBr$_6$} & \textit{Null}
    & Ti $3d_{xy}$ & Indirect $\Gamma\to$X & \textit{Null} 
    & Ti $3d_{xy}$ & $\Gamma$ & high $\Lambda$ & high $\Lambda$ & non-hydr.\\
    \rule{0pt}{4ex}
    \textbf{Cs$_2$SnBr$_6$} & \textit{Null}  
    & Sn $5s$ & Direct $\Gamma$ & \textit{Null} 
    & Sn $5s$ & $\Gamma$ & low $\Lambda$ & low $\Lambda$ & hydr.\\
    \midrule
    \bottomrule
\end{tabular}
}
\caption{Metal-orbital character of the band edges, nature of the band gap, effective mass anisotropy and nature of the exciton for the analysed materials and vacancy-ordered perovskites Cs$_2$TeBr$_6$, Cs$_2$TiBr$_6$, and Cs$_2$SnBr$_6$. Contributions from halogen atoms are omitted for clarity.
}
\label{tab:exciton-prediction}
\end{table*}

In conclusion, we performed a detailed first-principles study of the optoelectronic properties of a set of representative Cs$_2$BB$'$Cl$_6$ double perovskites and compared them with those of the single perovskite CsPbCl$_3$ with its known \textit{hydrogenic} exciton series. Chemical substitution at the B and B$'$ metal sites allows for the realization of a wide variety of electronic structure properties with significant orbital-dependent effects on the anisotropy of charge-carrier effective masses and dielectric screening. Our calculations show that the chemical heterogeneity inherently present in double perovskites only leads to \textit{non-hydrogenic} excitons for B and B$'$ site combinations that result in indirect band gaps and large effective mass anisotropies at the band edges. In these double perovskites, excitons are strongly delocalized in reciprocal space and thus experience the full spatial variation of the dielectric screening. We show that our understanding of excitons in halide double perovskites can be extended to vacancy-ordered perovskites with a single B-site. The nature of excitons in double perovskites can thus be predicted based on computationally efficient DFT calculations. With these insights our state-of-the-art $GW$+BSE calculations can provide a starting point for the development of tight-binding models for excitons, aid in the interpretation of experiments, and inspire further study of excited state properties of this complex quarternary family of materials.
\begin{suppinfo}\label{Supporting Information}
Methodological and computational details (including convergence plots), DFT and QP electronic band structures, absorption spectra, exciton diagrams, radial probability density, and variation of dielectric constant in reciprocal space. Structure files (cif) used in calculations.
\end{suppinfo}

\begin{acknowledgement}\label{Acknowledgement}
We acknowledge computing resources provided by the Dutch national supercomputing center Snellius supported by the SURF Cooperative and PRACE for awarding access to the Marconi100 supercomputer at CINECA, Italy. This work was partially supported by the Dutch Research Council (NWO) under grant number OCENW.M20.337. MRF and YC gratefully acknowledge access to additional computational resources used for some of these calculations, provided by the Extreme Science and Engineering Discovery Environment (now ACCESS) to the supercomputer Stampede2 at the Texas Advanced Computing Center (TACC) through the allocation TG-DMR190070. MRF acknowledges support from the UK Engineering and Physical Sciences Research Council (EPSRC), Grant EP/V010840/1.
\end{acknowledgement}

\providecommand{\latin}[1]{#1}
\makeatletter
\providecommand{\doi}
  {\begingroup\let\do\@makeother\dospecials
  \catcode`\{=1 \catcode`\}=2 \doi@aux}
\providecommand{\doi@aux}[1]{\endgroup\texttt{#1}}
\makeatother
\providecommand*\mcitethebibliography{\thebibliography}
\csname @ifundefined\endcsname{endmcitethebibliography}
  {\let\endmcitethebibliography\endthebibliography}{}

\end{document}